\documentclass[fleqn,usenatbib]{mnras}

\usepackage{newtxtext,newtxmath}

\usepackage[dvipsnames]{xcolor}

\usepackage[T1]{fontenc}
\usepackage{ae,aecompl}

\usepackage{graphicx}	

\def\gtorder{\mathrel{\raise.3ex\hbox{$>$}\mkern-14mu
     \lower0.6ex\hbox{$\sim$}}}
\def\ltorder{\mathrel{\raise.3ex\hbox{$<$}\mkern-14mu
     \lower0.6ex\hbox{$\sim$}}}

\newcommand{\beq}{\begin{equation}}
\newcommand{\eeq}{\end{equation}}
\newcommand{\ba}{\begin{eqnarray}}
\newcommand{\ea}{\end{eqnarray}}
\def\spose#1{\hbox to 0pt{#1\hss}}
\newcommand{\lta}{\mathrel{\spose{\lower 3pt\hbox{$\mathchar"218$}}
      \raise 2.0pt\hbox{$\mathchar"13C$}}}
\newcommand{\gta}{\mathrel{\spose{\lower 3pt\hbox{$\mathchar"218$}}
      \raise 2.0pt\hbox{$\mathchar"13E$}}}

\newcommand{\comments}[1]{} 

\title[Magnetic fields are black hole catalyst]{Magnetic fields catalyze massive black hole formation and growth}

\author[M. C. Begelman and J. Silk]{Mitchell C. Begelman$^{1,2}$\thanks{E-mail: mitch@jila.colorado.edu} and Joseph Silk$^{3,4,5}$\\
$^{1}$JILA, University of Colorado and National Institute of Standards and Technology, 440 UCB, Boulder, CO 80309-0440, USA \\
$^{2}$Department of Astrophysical and Planetary Sciences, 391 UCB, Boulder, CO 80309-0391, USA \\
$^{3}$Institut d'Astrophysique de Paris, Sorbonne Universit\'es, UPMC Univ. Paris 06 et CNRS, UMR 7095, F-75014, Paris, France\\
$^{4}$Department of  Physics \& Astronomy, The Johns Hopkins University, Baltimore, MD 21218, USA\\
$^{5}$Beecroft Institute of Particle Astrophysics and Cosmology, Department of Physics, University of Oxford,  Oxford OX1 3RH, UK
}

\date{Accepted 2023 August 29. Received 2023 August 28; in original form 2023 May 30}

\pubyear{2023}

\begin{document}
\label{firstpage}
\pagerange{\pageref{firstpage}--\pageref{lastpage}}
\maketitle

\begin{abstract}
Large-scale magnetic fields in the nuclear regions of protogalaxies can promote the formation and early growth of supermassive black holes (SMBHs) by direct collapse and magnetically boosted accretion. Turbulence associated with gravitational infall and star formation can drive the rms field strength toward equipartition with the mean gas kinetic energy; this field has a generic tendency to self-organize into large, coherent structures.  If the poloidal component of the field (relative to the rotational axis of a star-forming disc) becomes organized on scales $\lesssim r$ and attains an energy of order a few percent of the turbulent energy in the disc, then dynamo effects are expected to generate magnetic torques capable of increasing the inflow speed and thickening the disc.  The accretion flow can transport matter toward the center of mass at a rate adequate to create and grow a massive direct-collapse black hole (DCBH) seed and fuel the subsequent AGN at a high rate, without becoming gravitationally unstable. Fragmentation and star formation are thus suppressed and do not necessarily deplete the mass supply for the accretion flow, in contrast to prevailing models for growing and fueling SMBHs through disc accretion.   

\end{abstract}


\section{Introduction}\label{sec:Introduction}

The formation and growth of supermassive black holes (SMBHs) in galactic nuclei remains one of the main unsolved problems of extragalactic astrophysics.  To form rare $10^9 M_\odot$ black holes responsible for high-redshift quasars \citep{venemans2017,banados2018} requires an early start seeded by Pop III stars and/or sustained growth of more massive seeds formed later, with accretion rates near or exceeding the Eddington limit.  Theoretical studies of the former model exposes difficulties in obtaining both the requisite merger rates and growth of individual seeds by accretion, especially in the presence of vigorous star formation.  For the second class of models --- direct collapse of infalling gas to form a supermassive star-like object which then collapses to a large black-hole seed --- the principal worry is that high infall rates required, exceeding a few tenths of a solar mass per year, are difficult to arrange, requiring a convergence of low angular momentum, very low metallicity, and the suppression of H$_2$ cooling.  Simulations suggest that these conditions requires high halo accretion rates and proximity to a source of H$_2$--dissociating photons \citep{wise2019}, although the proximity issue may be resolved if AGN are preferentially located in dense concentrations of galaxies or protoclusters, as recent evidence suggests \citep{overzier2022,tozzi2022}.  Whether such clustering suffices, however, remains unclear.
 
The recent discovery of  high-redshift AGN candidates \citep{larson2023, yang2023,maiolino23} helps to constrain the history of SMBH formation, but does not remove the need for a sustained period of near- to super-Eddington accretion, whatever the class of seeds \citep{inayoshi2020}. 
 
The origin of galactic magnetic fields is an equally profound puzzle.  Here the problem is to amplify a tiny seed field, most likely formed under cosmological conditions by a battery process, by 10--15 orders of magnitude.  Both analytic arguments and numerical simulations suggest that this happens  at an exponential rate (with the e-folding time set by dynamical scales) due to a small-scale dynamo driven by the turbulence arising from a combination of gravitational collapse and stellar processes \citep{schleicher10,latif13,schober13,sharda21}.  Once the field strength approaches equipartition levels with gas energy, it needs to be organized on large scales, but in theory, this  seems to be possible as well.  Thus, it is at least plausible that young protogalaxies already have substantial, and well-organized, magnetic fields.  

In this Letter, we argue that the problems of SMBH formation and the early development of large-scale, equipartition-level magnetic fields in protogalaxies are linked.  We suggest that the formation of direct-collapse  black holes (DCBHs) is greatly stimulated by, and may require, the presence of a well-organized background magnetic field in the environment.  Interacting with a centrifugally supported disc in a protogalactic (or galactic) nucleus, a poloidal field of sufficient strength can trigger rapid, ``magnetically boosted'' accretion driven by large-scale magnetic torques.  Both the inflow velocities and the scale heights of such magnetically boosted discs can be much larger than those predicted by traditional disc models in which angular momentum transport relies on internal turbulent viscosity and the thickness is governed by gas pressure gradients. 

Because of their lower densities, such discs are less likely to be gravitationally unstable, and thus less likely to fragment and form stars.  As a result, there is no need for near-zero metallicity or the destruction of H$_2$, the two principal stumbling blocks for DCBH formation except under the special conditions described above \citep{inayoshi2020}.  The main requirement is that the magnetically-induced inflow rate be large enough compared to the rate at which star formation depletes the mass reservoir.  In \S \ref{sec:sec2}, we present a heuristic discussion --- based on scalings extracted from a simple, kinematic dynamo model --- of the conditions under which magnetically boosted accretion can occur.  Crucially, the onset of magnetically boosted inflow at some outer radius likely ensures the disc's ability to accumulate and retain magnetic flux from the environment, thus establishing and maintaining the boosted state at smaller radii as well.  

We briefly review recent models for the evolution of magnetic fields in galaxies in \S \ref{sec:sec3}, to assess whether the required field levels are plausible. In \S \ref{sec:sec4} we discuss the main properties of magnetically boosted discs traversing both the galactic nuclear potential and the gravitational sphere of influence of the central mass concentration (e.g., an already formed and growing black hole).  While such discs may settle into a star-forming state, this is not inevitable and we derive a condition for star formation to have a minor impact on the accretion flow. 
 A range of initial or outer boundary conditions could trigger the magnetically-boosted flows discussed here, including cold accretion via the cosmic web or merger-driven accumulation of large molecular gas reservoirs in post-starburst galactic nuclei. In the latter case stellar tidal disruptions could be common \citep{french2015}, providing an additional growth pathway for intermediate-mass black holes
  \citep{Stone2020}. In all cases a magnetically boosted disc would allow mass to be supplied to the central region at very large rates, more than adequate to produce large DCBH seeds at early times.  Additionally, magnetically boosted accretion provides an attractive mechanism for resolving the long-standing problem of fueling luminous AGN, which is challenging because of the predicted role of self-gravity in quasar disc models  \citep{kolykhalov80,shlosman87,shlosman89,goodman03}.  We summarize and discuss the proposed model in \S \ref{sec:sec5}.

\section{Magnetically boosted accretion}
\label{sec:sec2}  

Suppose a turbulent disc of gas settles into the central region of a galaxy, in the presence of a net vertical magnetic field $B_z$. We also assume that the disc suffers some angular momentum loss, which drives inflow at a speed $v_r$. If the vertical field is steady, $\partial B_z/\partial t = 0 $, then the induction equation can be written 
\begin{equation}
    \label{eq:dynamo1}
    \nabla \times \left[ {\bf v} \times {\bf B} - \eta (\nabla \times {\bf B})\right] = 0 \ , 
\end{equation}
where $\eta$ is some effective electrical resistivity, presumably the result of turbulence. A qualitative analysis of the $\phi$-component of equation (\ref{eq:dynamo1})  then implies that the field must develop a radial component of strength
\begin{equation}
    \label{br}
    B_r \sim {v_r H \over \eta} B_z \ ,
\end{equation}
where $H$ is the disc scale height.

In the presence of rotational shear,  a toroidal field builds up at a rate $dB_\phi/dt \sim \Omega B_r$, until it saturates after a vertical diffusion timescale $t_{\rm diff} \sim H^2/\eta$.  In equilibrium, $B_\phi$ is thus given by 
 \begin{equation}
    \label{bphi}
    B_\phi \sim  \Omega t_{\rm diff} B_r \sim {H^2 \Omega \over \eta} B_r \ ,
\end{equation}
where $\Omega$ is the Keplerian angular velocity in the external gravitational potential. 

As \cite{lubow94} pointed out, the magnetic field may not be in a steady state, because equation (\ref{br}) implies a specific ratio of radial to vertical field, $\xi \equiv B_r / B_z $, for given disc parameters $v_r$, $H$ and $\eta$, that may not be attainable.  Indeed, for inflow driven by turbulent viscosity with a \cite{shakura73} $\alpha$-parameter and turbulent magnetic Prandtl number $ \sim 1$, $v_r \sim H^2 \Omega / r$ and a steady state requires $\xi \sim H/r \ll 1$.  But $\xi$ is most likely determined by properties of the magnetic field external to the disc \citep[e.g.,][]{blandford82,konigl89,li95,lubow94,okuzumi14,takeuchi14}, which tend to favor $\xi \sim O(1)$. For $\xi \gg H/r$, \cite{lubow94} showed that the flux would diffuse out of the disc faster than it can be advected inward.  

The \cite{lubow94} argument is incomplete, however, because it does not take into account the effect of the large-scale magnetic field on $v_r$.  In the absence of any viscous torque, this field would drive a radial velocity 
 \begin{equation}
    \label{sketch3}
    v_r \sim { 1\over H \Omega} { B_\phi B_z \over 4\pi\rho} \ ,
\end{equation}
by extracting angular momentum from the disc vertically. If a steady state holds, this inflow speed is already of the same order as the viscous inflow speed when $\xi \gta H/r$ and greatly exceeds it when $\xi $  is much larger.  Moreover, comparing equations (\ref{br}) and (\ref{bphi}) we find that $\xi$ drops out of the condition for a steady state, which becomes $v_{{\rm A}z} = \eta / H$, where $v_{{\rm A}z} = (B_z^2 / 4\pi \rho)^{1/2}$ is the Alfv\'en speed associated with the vertical field strength.  Flux is advected inward if $v_{{\rm A}z} > \eta / H$ and escapes if the opposite holds, regardless of the value of $\xi$.  

To check whether the net vertical field can plausibly become strong enough to trigger flux advection, we note that $\eta$ is related to the turbulent pressure, which contributes to determining the disc scale height, $H$.  We can write $\eta \sim q u_t^2 \Omega^{-1}$ where $u_t$ is the typical velocity (or fluctuating Alfv\'en speed) in a turbulent cell, $\Omega^{-1}$ sets the correlation timescale and $q$ is a constant that could depend on a number of factors including the driving mechanism of the turbulence, the background magnetic field and shear, and gas thermodynamics.  The important point is that $q$ could be quite small.  For example, in the standard magnetorotational instability, empirical results from shearing-box simulations yield $q \sim 0.3$ Pr$_{\rm t}^{-1}$ \citep{salvesen16}, where Pr$_{\rm t}^{-1}$ is the turbulent Prandtl number, which could be as large as a few \citep{bian21}.   

Both the turbulent velocity and gas pressure contribute to determining the scale height of the disc.  The organized $B_\phi$ and $B_r$ components also affect the vertical structure, but in a disc threaded by net flux these components are antisymmetric about the midplane and therefore squeeze the disc rather than supporting it against gravity within the inner scale height. Neglecting this squeezing effect, we estimate the scale height to be $H \sim (u_t^2 + c_s^2)^{1/2} \Omega^{-1}$, where $c_s$ could also include radiation pressure support.  We can also define the scale height due to turbulence alone, $H_t \sim u_t \Omega^{-1} < H$. Using our parametrization for $\eta$ above, the condition for trapping and advecting ambient magnetic flux can be written    
\begin{equation}
    \label{sketch4}
    v_{{\rm A}z} \gtrsim  q {H_t \over H} u_t.
\end{equation}
We will argue in \S 3 that this condition may be attainable in galactic nuclei. 

If condition (\ref{sketch4}) holds at some outer radius, $R_{\rm out}$, then steady accretion boosted by magnetic torques can be maintained at smaller $r$ provided that the flow can establish the equilibrium condition $v_{{\rm A}z} \sim q (H_t / H) u_t$ through radial rearrangement of magnetic flux and gas density.  We suggest that this will happen automatically, since magnetic flux advected inward through $R_{\rm out}$ will be effectively trapped, unable to diffuse away.  We might imagine, for example, that the flux is effectively advected inward between $R_{\rm out}$ and some smaller radius, $R_1$, but decouples from the gas at still smaller radii. A layer of enhanced flux would then accumulate between $R_1$ and $R_{\rm out}$.  Ultimately the flux would leak out of this layer, but not entirely in the outward direction. We would expect flux leaking inward to gradually establish the flux-trapping condition at smaller and smaller radii (i.e., reducing the value of $R_1$) until the entire region within $R_{\rm out}$ is in a magnetically boosted accretion state. 

Magnetic trapping could be established ab initio if the disc density is regulated at a constant Toomre $Q$-parameter, e.g., close to the critical value for gravitational instability, $Q \gtrsim 1$ (possibly $\gtrsim 3$ if the disc is strongly magnetized: \citealt{riols16}).  We would expect this to be the case if the turbulence is maintained by continual star formation through stellar winds and supernovae. If we could ignore magnetic flux diffusion, the magnetic flux-to-mass ratio, $\Psi \equiv B_z/(\rho H)$, would be conserved.  Setting $H \sim H_t$ for a disc thickened primarily by turbulence, we can write \begin{equation}
    \label{sketch5}
    {v_{{\rm A}z} \over q u_t } \sim \Psi \left( {\rho \over \Omega^2} \right)^{1/2} \  .
\end{equation}
For $Q \approx$ const., the density tracks $\rho \propto \Omega^2$ as a function of radius.  We then have $v_{{\rm A}z}/q u_t \propto \Psi $, and conserving $\Psi$ then implies that the threshold condition for magnetic trapping continues to hold at $r < R_{\rm out}$.  

More generally, the disc might maintain its turbulent support without the aid of stellar processes, i.e., at $Q \gg 1$.  In this case the turbulence well inside $R_{\rm out}$ would most likely derive from the dynamo itself.  As an example of how this process might work, \cite{begelman2023} conjectured that adequate turbulence to support a disc vertically could be generated by tearing instabilities triggered by the vertical gradient of $B_\phi$.  Adopting their estimate of $\eta \sim 0.1 v_{{\rm A}\phi} H$ for this process gives $u_t \sim 0.1 q^{-1} v_{{\rm A}\phi}$ or, equivalently, $v_{{\rm A}\phi} \sim 10 v_{{\rm A}z}$, where $v_{{\rm A}\phi}$ is the Alfv\'en speed associated with the toroidal dynamo field.  We then have $\rho \propto \Omega^2 / \Psi^2$.  In general, we would expect $\rho$ to decrease with radius more gradually than $\propto \Omega^2$, particularly in the inner regions where a large accumulated mass (e.g., an SMBH) dominates the gravitational potential.  In this case, the flux per unit mass $\Psi$ would have to increase toward smaller radii, as a result of the diffusion process discussed above.

\section{Coherent magnetic fields in protogalaxies}
\label{sec:sec3}
 
Several lines of theoretical and observational analysis point to the plausibility not only that primordial magnetic fields can be amplified to near-equipartition strength early in galactic evolution, but also that these fields can become self-organized on system-size scales.  On the observational side, statistical evidence for Faraday rotation along lines of sight to quasars, modeling of chemical abundances affected by cosmic ray transport, and direct detection of field strength and geometry in high-redshift galaxies point to early growth and organization of galactic magnetic fields \citep{kulsrud08, mao17}. Theoretically, studies of small-scale turbulent dynamos show exponentially fast growth toward equipartition levels, on dynamical timescales \citep{kulsrud08}.  Dynamos driven by gravitationally-induced turbulence associated with infalling streams of gas and stellar processes such as supernovae and stellar winds can quickly lead to equipartition fields in the central regions of galaxies, even when starting from extremely small cosmological seed fields established by battery mechanisms or more exotic processes \citep{pudritz89,kulsrud97}.  Dynamos interior to massive stars and accreting black holes can also contribute to amplification of the small-scale field.  Once the small-scale fields are present, self-organization seems to follow naturally through inverse cascades of turbulent magnetic energy to large scales, both with \citep{frisch75} and without \citep{zrake14,brandenburg15} the presence of magnetic helicity.  

The Central Molecular Zone (CMZ) of the Milky Way, which occupies the inner few hundred parsecs, provides a test case for the existence of a highly organized field in a mature galactic nucleus. While the main body of molecular gas appears to form an inhomogeneous, turbulent disc pervaded by a toroidal field, the field on either side of the disc appears to be poloidal, with long, coherent filaments outlined by synchrotron emission \citep{yusefzadeh22}.  Although the 
field within the filaments may approach mG strengths \citep{yusefzadeh87,chandran00,morris06,morris15}, the mean field strength is uncertain \citep{ferriere11} and may be as small as $\sim 10 \mu$G \citep{larosa05,yusefzadeh13}. 
Given estimates of the mass and turbulent energy density of the gas in the CMZ, this would suggest that $v_{{\rm A}z}$ could be slightly below or close to the level needed to trigger magnetically boosted accretion.

A large body of theoretical work,  both analytic and computational, suggests that near-equipartition fields developed during the earliest epochs of galaxy formation, possibly during the burst of Population III star formation that preceded the era when DCBHs would have formed \citep{schleicher10,sharda21}.  Turbulence, an unavoidable byproduct of gravitational collapse \citep{greif08,federrath11}, leads to exponential growth of a seed field through a small-scale dynamo \citep{schleicher10,schober13,latif13}.  The subsequent development of a large-scale field has not been clearly demonstrated by simulations --- possibly due to their limited resolution and/or duration --- although this outcome is thought likely.

Thus, it does not seem much of an extrapolation to postulate that a well-organized poloidal field, with strength of order a few percent of the turbulent energy density, existed during the formative stages of protogalaxies.   Such a field might be maintained at high strength by the turbulent, cool gas in a star-forming disc.  But even at radii where this coupling is weak, the magnetic field could be maintained by turbulence induced by infalling streams from the cosmic web  at early times \citep{pudritz89}, and the multiphase, supernova-driven interstellar medium at later epochs \citep{mckee77,gent23}.  For the parameter $q \sim 0.1$, we see from  \S\ref{sec:sec2} that a poloidal field with energy density $\gtrsim 1 \%$ that of the turbulence should be adequate to trigger magnetically boosted accretion all the way in to the galactic center. In the following, we will assume that this holds and explore its consequences.

\section{Accretion vs. star formation}\label{sec:sec4}

According to the magnetically boosted accretion model, the inflow speed is $v_r \sim \xi v_{{\rm A}z}$ and the disc scale height satisfies $H/r \sim q^{-1} (v_{{\rm A}z}/ v_K)$, where $v_K$ is the Keplerian speed in the background potential; we will henceforth assume $\xi \sim 1$ and normalize $q$ to 0.1. We can parametrize the density as $\rho \sim v_K^2 / (2\pi G r^2 Q)$, where $Q$ is the Toomre parameter but we make no a priori assumptions about whether $Q$ is close to 1 (gravitationally unstable) or is much larger (stable).  The accretion rate is then given by  
\begin{equation}
    \label{mdot4}
    \dot M = 4\pi \rho v_r H r \sim 0.2 q_{-1} {  v_K^3  \over G Q } \left(  { H \over r }\right)^2  \ . 
\end{equation}
For $\dot M$ independent of radius, $H/r \propto Q^{1/2} v_K^{-3/2}$.  Modeling the galactic potential as an isothermal sphere with velocity dispersion $\sigma$, this implies $H/r \propto Q^{1/2}$, allowing for a uniform aspect ratio and $\rho \propto r^{-2}$ in a disc with constant $Q$, as discussed in \S \ref{sec:sec2} in connection with the advection of magnetic flux.  However, inward transport of matter will build up a central mass $M(t)$, which deepens the gravitational potential within a radius $r_B = GM/\sigma^2$. To within factors of order unity, the expressions describing the magnetically boosted accretion model can be extrapolated inward into this Keplerian region by replacing $\sigma$ by $v_K = (GM/r)^{1/2}$.  We then find $H/r \propto Q^{1/2} r^{3/4}$, indicating that the disc must become geometrically thinner and/or less self-gravitating (larger $Q$) toward smaller radii. 

We can characterize the star formation rate per unit disc surface area by 
\beq
\label{Sigmadot}
\dot \Sigma_* = \epsilon_* \Omega \Sigma \ ,
\eeq
where $\epsilon_*$ is the star formation efficiency and $\Sigma \sim 2\rho H$ is the surface density.  For the marginally self-gravitating case $Q \sim 1$, we expect $\epsilon_*$ to depend on feedback from stellar winds and supernovae, but generally to lie between $\sim 10^{-3}$ and a few percent \citep{begelman17}. Even for discs with a mean density satisfying $Q \gg 1$, there may be some fragmentation and star formation due to thermal instability leading to the formation of embedded molecular clouds, which could then form stars via the usual processes. Given this uncertainty, we will simply parametrize $\epsilon_*$ in units of $10^{-2}$. To within factors of order unity, the star formation rate out to a radius $r$ is given by   
\begin{equation}
    \label{mdot5}
    \dot M_* = \pi \epsilon_* \Sigma \Omega r^2 \sim 0.01 \epsilon_{-2} {  v_K^3  \over G Q } \left(  { H \over r }\right)  \ . 
\end{equation}
In order for most of the accretion mass flux to reach a radius $r$ without forming stars, we must have $\dot M \gtrsim \dot M_*$ or, equivalently,
\begin{equation}
    \label{mdot6}
    {H \over r} \gtrsim 0.05 {\epsilon_{-2} \over q_{-1}}   \ . 
\end{equation}
For the case where $H/r$ is regulated by star formation with $Q \sim 1$, the simple feedback argument presented in \cite{begelman17} (with dimensionless supernova rate and energy injection parameters set equal to 1) gives  
$u_t \gtrsim 20 \epsilon_{-2}^{1/2}$ km s$^{-1}$ (cf.~equation 30 of \citealt{begelman17}). For the outer region of the disc with constant $\sigma = 200 \sigma_{200}$ km s$^{-1}$, this gives
$H/r \gtrsim 0.1 \epsilon_{-2}^{1/2} \sigma_{200}^{-1} $, suggesting that star formation is unlikely to limit accretion in this outer region.  However, within the sphere of influence of the central mass (presumably a growing black hole), we saw above that $H/r \propto  r^{3/4}$, suggesting that star formation would rapidly become an important mass sink.  
For the case where the disc thickness is determined by turbulent magnetic pressure, it is hard to predict the value of $H/r$ --- it may depend on outer boundary conditions, e.g., how the matter infalling from the outer galaxy or cosmic web transitions to a disc.
For this case, it is reasonable to suppose that $H/r$ is roughly constant, implying that $Q \propto r^{-3/2}$ inside the sphere of influence, suppressing disc self-gravity at small radii.

The magnetically boosted accretion scenario thus provides a mechanism for growing a supermassive black hole, or fueling an AGN, at potentially any rate lower than the theoretical maximum set by the gravitational potential of the galactic nucleus, $\dot M_{\rm max} \sim \sigma^3/G \approx 1900 \sigma_{200}^3 M_\odot$ yr$^{-1}$. For our fiducial parameters, 
 \begin{equation}
    \label{mdot}
    \dot M \sim 380 {q_{-1} \sigma_{200}^3 \over Q} \left( {H \over r} \right)^2 \  {\rm M_\odot \ yr^{-1}}  \gtrsim  0.9  {\epsilon_{-2}^2 \sigma_{200}^3 \over q_{-1} Q}  \   {\rm M_\odot \ yr^{-1}},
\end{equation}
where the last relation assumes that star formation has little effect on the accretion rate.  Note that the actual mass supply to the outer disc sets $\dot M$, and thus determines the value of $Q^{-1} (H/r)^2$ in a steady state.

If we assume that the black hole is growing toward an eventual final mass $M_{\sigma} = 3 \times 10^8 \sigma_{200}^4$ M$_\odot$ according to the $M-\sigma$ relation, then the accretion rate is 
 \begin{equation}
    \label{mdot2}
    {\dot M \over \dot M_E}  \sim 46   {q_{-1}  \over Q} \left( {H \over r} \right)^2  \left( {M \over M_\sigma} \right)^{-4}  \sigma_{200}^{-1}  
\end{equation}
in  units of the Eddington accretion rate, $\dot M_E = 10 L_E/c^2$. Magnetically boosted accretion is thus capable of continuing to fuel the SMBH at close to the Eddington limit during its quasar phase.

In addition to the effects of magnetic pressure support and supernova feedback discussed here, simulations suggest that infalling clumps of gas will inevitably stir up and thicken the disc \citep{beckmann2019}. This could also help to reduce $\epsilon_*$ and increase $Q$, further reducing star formation.

\section{Discussion and conclusions}\label{sec:sec5}

We have proposed that the development of large-scale magnetic fields in the central regions of protogalaxies may be a crucial factor in stimulating the formation and early growth of SMBH seeds by direct collapse, and in the subsequent fueling of  AGN.  Our model, while replete with undetermined factors, indicates a pathway by which matter collecting in the central nuclear region can interact with the background field to create a magnetically boosted accretion flow, which avoids fragmentation due to self-gravity as it funnels mass into the center at rates that can reach super-Eddington values relative to DCBH seeds and even the most luminous quasars.  
Fragmentation and star formation can be suppressed because magnetically boosted accretion flows have much lower densities than found in standard disc models with  similar accretion rates.  
The principal constraints on previous DCBH models --- notably, extremely low metallicity and suppression of H$_2$ \citep{inayoshi2020} --- are obviated by our model.

Recent works have emphasized the possible role of magnetic fields in promoting the formation of DCBHs \citep{latif14,latif23a,latif23b}, mainly through enhanced transport of angular momentum and the suppression of fragmentation due to magnetic pressure effects. However, the simulations leading to these results do not treat the possible existence of an already organized field on large scales as has been suggested by many other simulations and analytic models that start from halo-scale conditions \citep{schleicher10,schober13,latif13,sharda21}.  Our argument is that these large-scale fields, and the torques they exert, can drive qualitatively faster inflow than that driven by the Maxwell and Reynolds stresses associated with small-scale turbulent fields.  The resulting lower densities in the inflow
further reduce the effects of self-gravity. Numerical experiments, initialized with suitable large-scale fields, should be able to test these predictions.

The key condition for triggering such a flow is that the energy density of the organized poloidal field at some outer radius exceed some fraction (nominally of order 1 percent) of the turbulent energy density.  Once triggered, such a flow would tend to perpetuate itself by trapping magnetic flux and advecting it inwards until it reaches a steady state.  The density is reduced both by faster inflow speeds --- the result of large-scale magnetic torques supplanting turbulent viscosity as the main angular momentum transport mechanism --- and by potentially larger scale heights supported by a combination of turbulent magnetic pressure and an organized toroidal field.   

The predicted large-scale torques can readily be compared to the internal ``viscous'' torques resulting from turbulence.  According to equations (\ref{bphi}) and (\ref{sketch3}), the inflow speed driven by large-scale magnetic torques is $v_r \sim \xi v_{{\rm A}z}^2 H/\eta $, whereas that driven by turbulent stress from any cause is $v_r \sim {\rm Pr_t} \eta / r$.  The condition for the large-scale stress to dominate the inflow rate is thus
\begin{equation}
    \label{sketch6}
    v_{{\rm A}z} > \left( {{\rm Pr_t} \over \xi} \right)^{1/2}  {\eta  \over \left( H r\right)^{1/2} }.
\end{equation}
In the usual limit of the magnetorotational instability (MRI), applicable when $v_{{\rm A}\phi}$ is smaller than the characteristic turbulent (or sound) speed, $\eta$ scales as $v_{{\rm A}z} H $ with a coefficient of a few, based on numerous local \citep{hawley95,bai13,salvesen16} and global \citep{mishra20} simulations of thermally supported discs.\footnote{Note that these torques are dominated by the Maxwell stress, which is several times larger than the Reynolds stress.} Therefore, the large-scale torques will dominate whenever $\xi = B_r/B_z$ exceeds a few times $H/r$. Turbulence driven by MRI is likely to be weaker in discs where $v_{{\rm A}\phi} > c_s$ \citep{begelman2023}, further relaxing the condition for the dominance of large-scale torques.   

Kinematic and gravitational effects can also lead to the rapid accumulation of gas in protogalactic halos.  In rare cases where cold streams feed gas with unusually low angular momentum into the center, angular momentum transport may not be necessary at all \citep{latif22}.  In the more general case with a large enough angular momentum to circularize the flow, infall can proceed at nearly the free-fall rate via global gravitational torques through the ``bars within bars'' mechanism \citep{shlosman89b}.  Such self-gravitating discs are likely to be subject to local gravitational instability leading to fragmentation, however, unless an adequate level of turbulence can be maintained \citep{begelman09}.  Moreover, the bars-within-bars mechanism is suppressed once the accumulated mass within a given radius becomes comparable to the disc mass.  Once this happens, further inflow due to large-scale gravitational torques can be maintained by the development of an eccentric ($m=1$) disc, albeit under somewhat less robust conditions \citep{hopkins10}.  Alternatively, ``gravitoturbulence'' on small scales can drive inflow at a rate given by the viscous formula with viscosity parameter $\alpha \lesssim 0.13$ \citep{gammie01}. Setting $\eta = \alpha c_s H/ {\rm Pr_t}$ and applying condition (\ref{sketch6}), we find that the large-scale torque will dominate the angular momentum transport whenever the vertical-flux plasma parameter $\beta_z = (c_s/v_{{\rm A}z})^2$ satisfies
\begin{equation}
    \label{sketch7}
    \beta_z <   {{\rm Pr_t} \xi \over \alpha^2}   {r  \over  H },
\end{equation}
i.e., even for very weak vertical fields. Moreover, in order for gravitoturbulence to operate without widespread fragmentation the cooling time scale must exceed $3 \Omega^{-1}$; this condition is unlikely to be satisfied under most conditions in a protogalaxy or in the outer disc fueling a luminous AGN. We note, however, that the operation of gravitational instabilities in the presence of strong MRI is still poorly understood \citep{riols18,riols19}; with much uncertainty about the relative and absolute importance of Reynolds, Maxwell and gravitational stresses in the resulting turbulence.

We have derived conditions necessary to minimize the impact of fragmentation and star formation on the mass flux reaching the inner disc, but the model does not intrinsically exclude star formation as an important (or even dominant) channel for the mass flow.  The dynamo underlying the boosted accretion model relies on a balance between amplification of the organized toroidal magnetic field due to Keplerian shear and its limitation due to turbulent resistivity.  The origin of the turbulence causing the resistivity is not specified by the generic model.  In cases where star formation is minimized, the turbulence could be triggered directly by the shear, as in the MRI invoked in standard accretion disc  models \citep{balbus1998}, or could involve other instabilities such as tearing modes exploiting the vertical gradient of $B_\phi$ \citep{begelman2023}.\footnote{Note that the source of energy for the tearing mode is still ultimately the shear, which is responsible for building up  $B_\phi$ in the first place.}  Alternatively, magnetically boosted inflows could be supported by  turbulence driven by supernovae, in a star-forming disc hovering close to marginal gravitational instability (Toomre parameter $Q \sim 1$).  In this case, star formation could be sufficiently abundant to regulate how much mass reaches the center. High-resolution molecular gas observations can play a key role in testing our model. 

In either limit, magnetic coupling is the crucial component of our model. The configuration adopted by the disc, including its scale height and inflow speed, could depend on outer boundary conditions where the mass supply first circularizes.   Even in cases where magnetic pressure support maintains the disc at $Q \gg 1$, molecular clouds might condense out of the  background disc and produce stars at a significant rate.   Higher inflow speeds, compared to viscous models, couple with magnetic levitation to inhibit any tendency to run-away fragmentation and excessive star formation.  Boosted accretion enables gas inflow at Alfv\'enic velocities and allows super-Eddington black-hole growth at early epochs and near-Eddington accretion at later times. 
 
A compelling aspect of the magnetically boosted accretion scenario is that it simultaneously explains how quasars can exist at all.  It has long been understood that accretion discs feeding luminous AGN should become gravitationally unstable and fragment if they extend further than $\sim 1$ pc from the central black hole   \citep{kolykhalov80,shlosman87,shlosman89,goodman03}.  One of the early motivations for studying magnetically elevated accretion discs --- discs that are thickened vertically by magnetic pressure but otherwise behave as normal viscous discs --- was to address the quasar self-gravity problem \citep{pariev03,begelman07,gaburov12}. The densities of magnetically boosted discs are lower than viscous discs of the same scale-height by the ratio of inflow speeds, $\sim \alpha^{-1} (r/H)$.  Since their heights may also be inflated by magnetic pressure, they potentially provide an even more robust resolution for longstanding theoretical challenges to growing and fueling SMBHs in galactic nuclei.    
          
\section*{Acknowledgements}
We thank the referee for thoughtful comments and critiques that helped us to improve the paper considerably.  MCB acknowledges support from NASA Astrophysics Theory Program grants NNX17AK55G and 80NSSC22K0826
and NSF grant AST 1903335, and thanks the Institut d'Astrophysique de Paris and the Institut Lagrange de Paris for their hospitality and support.
JS was supported in part 
by ERC Project No. 267117 (DARK) hosted by Universit\'e Pierre et Marie Curie  (UPMC), Paris 6. JS also acknowledges the support of the JHU by NSF grant OIA 1124403.
We thank Phil Armitage for helpful discussions.

\section*{Data Availability}
No new data was generated or analyzed to support the work in this paper.



\bibliographystyle{mnras}
\bibliography{biblio} 

\begin{thebibliography}{}
\makeatletter
\relax
\def\mn@urlcharsother{\let\do\@makeother \do\$\do\&\do\#\do\^\do\_\do\%\do\~}
\def\mn@doi{\begingroup\mn@urlcharsother \@ifnextchar [ {\mn@doi@}
  {\mn@doi@[]}}
\def\mn@doi@[#1]#2{\def\@tempa{#1}\ifx\@tempa\@empty \href
  {http://dx.doi.org/#2} {doi:#2}\else \href {http://dx.doi.org/#2} {#1}\fi
  \endgroup}
\def\mn@eprint#1#2{\mn@eprint@#1:#2::\@nil}
\def\mn@eprint@arXiv#1{\href {http://arxiv.org/abs/#1} {{\tt arXiv:#1}}}
\def\mn@eprint@dblp#1{\href {http://dblp.uni-trier.de/rec/bibtex/#1.xml}
  {dblp:#1}}
\def\mn@eprint@#1:#2:#3:#4\@nil{\def\@tempa {#1}\def\@tempb {#2}\def\@tempc
  {#3}\ifx \@tempc \@empty \let \@tempc \@tempb \let \@tempb \@tempa \fi \ifx
  \@tempb \@empty \def\@tempb {arXiv}\fi \@ifundefined
  {mn@eprint@\@tempb}{\@tempb:\@tempc}{\expandafter \expandafter \csname
  mn@eprint@\@tempb\endcsname \expandafter{\@tempc}}}

\bibitem[\protect\citeauthoryear{{Ba{\~n}ados} et~al.,}{{Ba{\~n}ados}
  et~al.}{2018}]{banados2018}
{Ba{\~n}ados} E.,  et~al., 2018, \mn@doi [\nat] {10.1038/nature25180}, \href
  {https://ui.adsabs.harvard.edu/abs/2018Natur.553..473B} {553, 473}

\bibitem[\protect\citeauthoryear{{Bai} \& {Stone}}{{Bai} \&
  {Stone}}{2013}]{bai13}
{Bai} X.-N.,  {Stone} J.~M.,  2013, \mn@doi [\apj]
  {10.1088/0004-637X/767/1/30}, \href
  {https://ui.adsabs.harvard.edu/abs/2013ApJ...767...30B} {767, 30}

\bibitem[\protect\citeauthoryear{{Balbus} \& {Hawley}}{{Balbus} \&
  {Hawley}}{1998}]{balbus1998}
{Balbus} S.~A.,  {Hawley} J.~F.,  1998, \mn@doi [Reviews of Modern Physics]
  {10.1103/RevModPhys.70.1}, \href
  {https://ui.adsabs.harvard.edu/abs/1998RvMP...70....1B} {70, 1}

\bibitem[\protect\citeauthoryear{{Beckmann}, {Devriendt}  \& {Slyz}}{{Beckmann}
  et~al.}{2019}]{beckmann2019}
{Beckmann} R.~S.,  {Devriendt} J.,   {Slyz} A.,  2019, \mn@doi [\mnras]
  {10.1093/mnras/sty2890}, \href
  {https://ui.adsabs.harvard.edu/abs/2019MNRAS.483.3488B} {483, 3488}

\bibitem[\protect\citeauthoryear{{Begelman} \& {Armitage}}{{Begelman} \&
  {Armitage}}{2023}]{begelman2023}
{Begelman} M.~C.,  {Armitage} P.~J.,  2023, \mn@doi [\mnras]
  {10.1093/mnras/stad914}, \href
  {https://ui.adsabs.harvard.edu/abs/2023MNRAS.521.5952B} {521, 5952}

\bibitem[\protect\citeauthoryear{{Begelman} \& {Pringle}}{{Begelman} \&
  {Pringle}}{2007}]{begelman07}
{Begelman} M.~C.,  {Pringle} J.~E.,  2007, \mn@doi [\mnras]
  {10.1111/j.1365-2966.2006.11372.x}, \href
  {https://ui.adsabs.harvard.edu/abs/2007MNRAS.375.1070B} {375, 1070}

\bibitem[\protect\citeauthoryear{{Begelman} \& {Shlosman}}{{Begelman} \&
  {Shlosman}}{2009}]{begelman09}
{Begelman} M.~C.,  {Shlosman} I.,  2009, \mn@doi [\apjl]
  {10.1088/0004-637X/702/1/L5}, \href
  {https://ui.adsabs.harvard.edu/abs/2009ApJ...702L...5B} {702, L5}

\bibitem[\protect\citeauthoryear{{Begelman} \& {Silk}}{{Begelman} \&
  {Silk}}{2017}]{begelman17}
{Begelman} M.~C.,  {Silk} J.,  2017, \mn@doi [\mnras] {10.1093/mnras/stw2533},
  \href {https://ui.adsabs.harvard.edu/abs/2017MNRAS.464.2311B} {464, 2311}

\bibitem[\protect\citeauthoryear{{Bian}, {Shang}, {Blackman}, {Collins}  \&
  {Aluie}}{{Bian} et~al.}{2021}]{bian21}
{Bian} X.,  {Shang} J.~K.,  {Blackman} E.~G.,  {Collins} G.~W.,   {Aluie} H.,
  2021, \mn@doi [\apjl] {10.3847/2041-8213/ac0fe5}, \href
  {https://ui.adsabs.harvard.edu/abs/2021ApJ...917L...3B} {917, L3}

\bibitem[\protect\citeauthoryear{{Blandford} \& {Payne}}{{Blandford} \&
  {Payne}}{1982}]{blandford82}
{Blandford} R.~D.,  {Payne} D.~G.,  1982, \mn@doi [\mnras]
  {10.1093/mnras/199.4.883}, \href
  {https://ui.adsabs.harvard.edu/abs/1982MNRAS.199..883B} {199, 883}

\bibitem[\protect\citeauthoryear{{Brandenburg}, {Kahniashvili}  \&
  {Tevzadze}}{{Brandenburg} et~al.}{2015}]{brandenburg15}
{Brandenburg} A.,  {Kahniashvili} T.,   {Tevzadze} A.~G.,  2015, \mn@doi [\prl]
  {10.1103/PhysRevLett.114.075001}, \href
  {https://ui.adsabs.harvard.edu/abs/2015PhRvL.114g5001B} {114, 075001}

\bibitem[\protect\citeauthoryear{{Chandran}, {Cowley}  \& {Morris}}{{Chandran}
  et~al.}{2000}]{chandran00}
{Chandran} B. D.~G.,  {Cowley} S.~C.,   {Morris} M.,  2000, \mn@doi [\apj]
  {10.1086/308184}, \href
  {https://ui.adsabs.harvard.edu/abs/2000ApJ...528..723C} {528, 723}

\bibitem[\protect\citeauthoryear{{Federrath}, {Sur}, {Schleicher}, {Banerjee}
  \& {Klessen}}{{Federrath} et~al.}{2011}]{federrath11}
{Federrath} C.,  {Sur} S.,  {Schleicher} D. R.~G.,  {Banerjee} R.,   {Klessen}
  R.~S.,  2011, \mn@doi [\apj] {10.1088/0004-637X/731/1/62}, \href
  {https://ui.adsabs.harvard.edu/abs/2011ApJ...731...62F} {731, 62}

\bibitem[\protect\citeauthoryear{{Ferri{\`e}re}}{{Ferri{\`e}re}}{2011}]{ferriere11}
{Ferri{\`e}re} K.,  2011, in {Morris} M.~R.,  {Wang} Q.~D.,   {Yuan} F.,  eds,
  Astronomical Society of the Pacific Conference Series Vol. 439, The Galactic
  Center: a Window to the Nuclear Environment of Disk Galaxies. p.~39

\bibitem[\protect\citeauthoryear{{French}, {Yang}, {Zabludoff}, {Narayanan},
  {Shirley}, {Walter}, {Smith}  \& {Tremonti}}{{French}
  et~al.}{2015}]{french2015}
{French} K.~D.,  {Yang} Y.,  {Zabludoff} A.,  {Narayanan} D.,  {Shirley} Y.,
  {Walter} F.,  {Smith} J.-D.,   {Tremonti} C.~A.,  2015, \mn@doi [\apj]
  {10.1088/0004-637X/801/1/1}, \href
  {https://ui.adsabs.harvard.edu/abs/2015ApJ...801....1F} {801, 1}

\bibitem[\protect\citeauthoryear{{Frisch}, {Pouquet}, {Leorat}  \&
  {Mazure}}{{Frisch} et~al.}{1975}]{frisch75}
{Frisch} U.,  {Pouquet} A.,  {Leorat} J.,   {Mazure} A.,  1975, \mn@doi
  [Journal of Fluid Mechanics] {10.1017/S002211207500122X}, \href
  {https://ui.adsabs.harvard.edu/abs/1975JFM....68..769F} {68, 769}

\bibitem[\protect\citeauthoryear{{Gaburov}, {Johansen}  \& {Levin}}{{Gaburov}
  et~al.}{2012}]{gaburov12}
{Gaburov} E.,  {Johansen} A.,   {Levin} Y.,  2012, \mn@doi [\apj]
  {10.1088/0004-637X/758/2/103}, \href
  {https://ui.adsabs.harvard.edu/abs/2012ApJ...758..103G} {758, 103}

\bibitem[\protect\citeauthoryear{{Gammie}}{{Gammie}}{2001}]{gammie01}
{Gammie} C.~F.,  2001, \mn@doi [\apj] {10.1086/320631}, \href
  {https://ui.adsabs.harvard.edu/abs/2001ApJ...553..174G} {553, 174}

\bibitem[\protect\citeauthoryear{{Gent}, {Mac Low}, {Korpi-Lagg}  \&
  {Singh}}{{Gent} et~al.}{2023}]{gent23}
{Gent} F.~A.,  {Mac Low} M.-M.,  {Korpi-Lagg} M.~J.,   {Singh} N.~K.,  2023,
  \mn@doi [\apj] {10.3847/1538-4357/acac20}, \href
  {https://ui.adsabs.harvard.edu/abs/2023ApJ...943..176G} {943, 176}

\bibitem[\protect\citeauthoryear{{Goodman}}{{Goodman}}{2003}]{goodman03}
{Goodman} J.,  2003, \mn@doi [\mnras] {10.1046/j.1365-8711.2003.06241.x}, \href
  {https://ui.adsabs.harvard.edu/abs/2003MNRAS.339..937G} {339, 937}

\bibitem[\protect\citeauthoryear{{Greif}, {Johnson}, {Klessen}  \&
  {Bromm}}{{Greif} et~al.}{2008}]{greif08}
{Greif} T.~H.,  {Johnson} J.~L.,  {Klessen} R.~S.,   {Bromm} V.,  2008, \mn@doi
  [\mnras] {10.1111/j.1365-2966.2008.13326.x}, \href
  {https://ui.adsabs.harvard.edu/abs/2008MNRAS.387.1021G} {387, 1021}

\bibitem[\protect\citeauthoryear{{Hawley}, {Gammie}  \& {Balbus}}{{Hawley}
  et~al.}{1995}]{hawley95}
{Hawley} J.~F.,  {Gammie} C.~F.,   {Balbus} S.~A.,  1995, \apj, 440, 742

\bibitem[\protect\citeauthoryear{{Hopkins} \& {Quataert}}{{Hopkins} \&
  {Quataert}}{2010}]{hopkins10}
{Hopkins} P.~F.,  {Quataert} E.,  2010, \mn@doi [\mnras]
  {10.1111/j.1365-2966.2010.17064.x}, \href
  {https://ui.adsabs.harvard.edu/abs/2010MNRAS.407.1529H} {407, 1529}

\bibitem[\protect\citeauthoryear{{Inayoshi}, {Visbal}  \& {Haiman}}{{Inayoshi}
  et~al.}{2020}]{inayoshi2020}
{Inayoshi} K.,  {Visbal} E.,   {Haiman} Z.,  2020, \mn@doi [\araa]
  {10.1146/annurev-astro-120419-014455}, \href
  {https://ui.adsabs.harvard.edu/abs/2020ARA&A..58...27I} {58, 27}

\bibitem[\protect\citeauthoryear{{Kolykhalov} \& {Sunyaev}}{{Kolykhalov} \&
  {Sunyaev}}{1980}]{kolykhalov80}
{Kolykhalov} P.~I.,  {Sunyaev} R.~A.,  1980, {Soviet Astronomy Letters}, 357,
  357

\bibitem[\protect\citeauthoryear{{K\"onigl}}{{K\"onigl}}{1989}]{konigl89}
{K\"onigl} A.,  1989, \mn@doi [\apj] {10.1086/167585}, \href
  {https://ui.adsabs.harvard.edu/abs/1989ApJ...342..208K} {342, 208}

\bibitem[\protect\citeauthoryear{{Kulsrud} \& {Zweibel}}{{Kulsrud} \&
  {Zweibel}}{2008}]{kulsrud08}
{Kulsrud} R.~M.,  {Zweibel} E.~G.,  2008, \mn@doi [Reports on Progress in
  Physics] {10.1088/0034-4885/71/4/046901}, \href
  {https://ui.adsabs.harvard.edu/abs/2008RPPh...71d6901K} {71, 046901}

\bibitem[\protect\citeauthoryear{{Kulsrud}, {Cen}, {Ostriker}  \&
  {Ryu}}{{Kulsrud} et~al.}{1997}]{kulsrud97}
{Kulsrud} R.~M.,  {Cen} R.,  {Ostriker} J.~P.,   {Ryu} D.,  1997, \mn@doi
  [\apj] {10.1086/303987}, \href
  {https://ui.adsabs.harvard.edu/abs/1997ApJ...480..481K} {480, 481}

\bibitem[\protect\citeauthoryear{{LaRosa}, {Brogan}, {Shore}, {Lazio}, {Kassim}
   \& {Nord}}{{LaRosa} et~al.}{2005}]{larosa05}
{LaRosa} T.~N.,  {Brogan} C.~L.,  {Shore} S.~N.,  {Lazio} T.~J.,  {Kassim}
  N.~E.,   {Nord} M.~E.,  2005, \mn@doi [\apjl] {10.1086/431647}, \href
  {https://ui.adsabs.harvard.edu/abs/2005ApJ...626L..23L} {626, L23}

\bibitem[\protect\citeauthoryear{{Larson} et~al.,}{{Larson}
  et~al.}{2023}]{larson2023}
{Larson} R.~L.,  et~al., 2023, \mn@doi [\apjl] {10.3847/2041-8213/ace619},
  \href {https://ui.adsabs.harvard.edu/abs/2023ApJ...953L..29L} {953, L29}

\bibitem[\protect\citeauthoryear{{Latif} \& {Schleicher}}{{Latif} \&
  {Schleicher}}{2023}]{latif23b}
{Latif} M.~A.,  {Schleicher} D. R.~G.,  2023, \mn@doi [\apjl]
  {10.3847/2041-8213/ace34f}, \href
  {https://ui.adsabs.harvard.edu/abs/2023ApJ...952L...9L} {952, L9}

\bibitem[\protect\citeauthoryear{{Latif}, {Schleicher}, {Schmidt}  \&
  {Niemeyer}}{{Latif} et~al.}{2013}]{latif13}
{Latif} M.~A.,  {Schleicher} D.~R.~G.,  {Schmidt} W.,   {Niemeyer} J.,  2013,
  \mn@doi [\mnras] {10.1093/mnras/stt503}, \href
  {https://ui.adsabs.harvard.edu/abs/2013MNRAS.432..668L} {432, 668}

\bibitem[\protect\citeauthoryear{{Latif}, {Schleicher}  \& {Schmidt}}{{Latif}
  et~al.}{2014}]{latif14}
{Latif} M.~A.,  {Schleicher} D.~R.~G.,   {Schmidt} W.,  2014, \mn@doi [\mnras]
  {10.1093/mnras/stu357}, \href
  {https://ui.adsabs.harvard.edu/abs/2014MNRAS.440.1551L} {440, 1551}

\bibitem[\protect\citeauthoryear{{Latif}, {Whalen}, {Khochfar}, {Herrington}
  \& {Woods}}{{Latif} et~al.}{2022}]{latif22}
{Latif} M.~A.,  {Whalen} D.~J.,  {Khochfar} S.,  {Herrington} N.~P.,   {Woods}
  T.~E.,  2022, \mn@doi [\nat] {10.1038/s41586-022-04813-y}, \href
  {https://ui.adsabs.harvard.edu/abs/2022Natur.607...48L} {607, 48}

\bibitem[\protect\citeauthoryear{{Latif}, {Schleicher}  \& {Khochfar}}{{Latif}
  et~al.}{2023}]{latif23a}
{Latif} M.~A.,  {Schleicher} D. R.~G.,   {Khochfar} S.,  2023, \mn@doi [\apj]
  {10.3847/1538-4357/acbcc2}, \href
  {https://ui.adsabs.harvard.edu/abs/2023ApJ...945..137L} {945, 137}

\bibitem[\protect\citeauthoryear{{Li}}{{Li}}{1995}]{li95}
{Li} Z.-Y.,  1995, \mn@doi [\apj] {10.1086/175657}, \href
  {https://ui.adsabs.harvard.edu/abs/1995ApJ...444..848L} {444, 848}

\bibitem[\protect\citeauthoryear{{Lubow}, {Papaloizou}  \& {Pringle}}{{Lubow}
  et~al.}{1994}]{lubow94}
{Lubow} S.~H.,  {Papaloizou} J.~C.~B.,   {Pringle} J.~E.,  1994, \mn@doi
  [\mnras] {10.1093/mnras/267.2.235}, \href
  {https://ui.adsabs.harvard.edu/abs/1994MNRAS.267..235L} {267, 235}

\bibitem[\protect\citeauthoryear{{Maiolino} et~al.,}{{Maiolino}
  et~al.}{2023}]{maiolino23}
{Maiolino} R.,  et~al., 2023, \mn@doi [arXiv e-prints]
  {10.48550/arXiv.2305.12492}, \href
  {https://ui.adsabs.harvard.edu/abs/2023arXiv230512492M} {p. arXiv:2305.12492}

\bibitem[\protect\citeauthoryear{{Mao} et~al.,}{{Mao} et~al.}{2017}]{mao17}
{Mao} S.~A.,  et~al., 2017, \mn@doi [Nature Astronomy]
  {10.1038/s41550-017-0218-x}, \href
  {https://ui.adsabs.harvard.edu/abs/2017NatAs...1..621M} {1, 621}

\bibitem[\protect\citeauthoryear{{McKee} \& {Ostriker}}{{McKee} \&
  {Ostriker}}{1977}]{mckee77}
{McKee} C.~F.,  {Ostriker} J.~P.,  1977, \mn@doi [\apj] {10.1086/155667}, \href
  {https://ui.adsabs.harvard.edu/abs/1977ApJ...218..148M} {218, 148}

\bibitem[\protect\citeauthoryear{{Mishra}, {Begelman}, {Armitage}  \&
  {Simon}}{{Mishra} et~al.}{2020}]{mishra20}
{Mishra} B.,  {Begelman} M.~C.,  {Armitage} P.~J.,   {Simon} J.~B.,  2020,
  \mn@doi [\mnras] {10.1093/mnras/stz3572}, \href
  {https://ui.adsabs.harvard.edu/abs/2020MNRAS.492.1855M} {492, 1855}

\bibitem[\protect\citeauthoryear{{Morris}}{{Morris}}{2006}]{morris06}
{Morris} M.,  2006, in Journal of Physics Conference Series. pp~1--9,
  \mn@doi{10.1088/1742-6596/54/1/001}

\bibitem[\protect\citeauthoryear{{Morris}}{{Morris}}{2015}]{morris15}
{Morris} M.~R.,  2015, in , Lessons from the Local Group: A Conference in honor
  of David Block and Bruce Elmegreen.
p.~391, \mn@doi{10.1007/978-3-319-10614-4_32}

\bibitem[\protect\citeauthoryear{{Okuzumi}, {Takeuchi}  \& {Muto}}{{Okuzumi}
  et~al.}{2014}]{okuzumi14}
{Okuzumi} S.,  {Takeuchi} T.,   {Muto} T.,  2014, \mn@doi [\apj]
  {10.1088/0004-637X/785/2/127}, \href
  {https://ui.adsabs.harvard.edu/abs/2014ApJ...785..127O} {785, 127}

\bibitem[\protect\citeauthoryear{{Overzier}}{{Overzier}}{2022}]{overzier2022}
{Overzier} R.~A.,  2022, \mn@doi [\apj] {10.3847/1538-4357/ac448c}, \href
  {https://ui.adsabs.harvard.edu/abs/2022ApJ...926..114O} {926, 114}

\bibitem[\protect\citeauthoryear{{Pariev}, {Blackman}  \& {Boldyrev}}{{Pariev}
  et~al.}{2003}]{pariev03}
{Pariev} V.~I.,  {Blackman} E.~G.,   {Boldyrev} S.~A.,  2003, \mn@doi [\aap]
  {10.1051/0004-6361:20030868}, \href
  {https://ui.adsabs.harvard.edu/abs/2003A&A...407..403P} {407, 403}

\bibitem[\protect\citeauthoryear{{Pudritz} \& {Silk}}{{Pudritz} \&
  {Silk}}{1989}]{pudritz89}
{Pudritz} R.~E.,  {Silk} J.,  1989, \mn@doi [\apj] {10.1086/167625}, \href
  {https://ui.adsabs.harvard.edu/abs/1989ApJ...342..650P} {342, 650}

\bibitem[\protect\citeauthoryear{{Riols} \& {Latter}}{{Riols} \&
  {Latter}}{2016}]{riols16}
{Riols} A.,  {Latter} H.,  2016, \mn@doi [\mnras] {10.1093/mnras/stw1112},
  \href {https://ui.adsabs.harvard.edu/abs/2016MNRAS.460.2223R} {460, 2223}

\bibitem[\protect\citeauthoryear{{Riols} \& {Latter}}{{Riols} \&
  {Latter}}{2018}]{riols18}
{Riols} A.,  {Latter} H.,  2018, \mn@doi [\mnras] {10.1093/mnras/stx2455},
  \href {https://ui.adsabs.harvard.edu/abs/2018MNRAS.474.2212R} {474, 2212}

\bibitem[\protect\citeauthoryear{{Riols} \& {Latter}}{{Riols} \&
  {Latter}}{2019}]{riols19}
{Riols} A.,  {Latter} H.,  2019, \mn@doi [\mnras] {10.1093/mnras/sty2804},
  \href {https://ui.adsabs.harvard.edu/abs/2019MNRAS.482.3989R} {482, 3989}

\bibitem[\protect\citeauthoryear{{Salvesen}, {Simon}, {Armitage}  \&
  {Begelman}}{{Salvesen} et~al.}{2016}]{salvesen16}
{Salvesen} G.,  {Simon} J.~B.,  {Armitage} P.~J.,   {Begelman} M.~C.,  2016,
  \mn@doi [\mnras] {10.1093/mnras/stw029}, \href
  {https://ui.adsabs.harvard.edu/abs/2016MNRAS.457..857S} {457, 857}

\bibitem[\protect\citeauthoryear{{Schleicher}, {Banerjee}, {Sur}, {Arshakian},
  {Klessen}, {Beck}  \& {Spaans}}{{Schleicher} et~al.}{2010}]{schleicher10}
{Schleicher} D.~R.~G.,  {Banerjee} R.,  {Sur} S.,  {Arshakian} T.~G.,
  {Klessen} R.~S.,  {Beck} R.,   {Spaans} M.,  2010, \mn@doi [\aap]
  {10.1051/0004-6361/201015184}, \href
  {https://ui.adsabs.harvard.edu/abs/2010A&A...522A.115S} {522, A115}

\bibitem[\protect\citeauthoryear{{Schober}, {Schleicher}  \&
  {Klessen}}{{Schober} et~al.}{2013}]{schober13}
{Schober} J.,  {Schleicher} D.~R.~G.,   {Klessen} R.~S.,  2013, \mn@doi [\aap]
  {10.1051/0004-6361/201322185}, \href
  {https://ui.adsabs.harvard.edu/abs/2013A&A...560A..87S} {560, A87}

\bibitem[\protect\citeauthoryear{{Shakura} \& {Sunyaev}}{{Shakura} \&
  {Sunyaev}}{1973}]{shakura73}
{Shakura} N.~I.,  {Sunyaev} R.~A.,  1973, \aap, \href
  {https://ui.adsabs.harvard.edu/abs/1973A&A....24..337S} {24, 337}

\bibitem[\protect\citeauthoryear{{Sharda}, {Federrath}, {Krumholz}  \&
  {Schleicher}}{{Sharda} et~al.}{2021}]{sharda21}
{Sharda} P.,  {Federrath} C.,  {Krumholz} M.~R.,   {Schleicher} D. R.~G.,
  2021, \mn@doi [\mnras] {10.1093/mnras/stab531}, \href
  {https://ui.adsabs.harvard.edu/abs/2021MNRAS.503.2014S} {503, 2014}

\bibitem[\protect\citeauthoryear{{Shlosman} \& {Begelman}}{{Shlosman} \&
  {Begelman}}{1987}]{shlosman87}
{Shlosman} I.,  {Begelman} M.~C.,  1987, \mn@doi [\nat] {10.1038/329810a0},
  \href {https://ui.adsabs.harvard.edu/abs/1987Natur.329..810S} {329, 810}

\bibitem[\protect\citeauthoryear{{Shlosman} \& {Begelman}}{{Shlosman} \&
  {Begelman}}{1989}]{shlosman89}
{Shlosman} I.,  {Begelman} M.~C.,  1989, \mn@doi [\apj] {10.1086/167526}, \href
  {https://ui.adsabs.harvard.edu/abs/1989ApJ...341..685S} {341, 685}

\bibitem[\protect\citeauthoryear{{Shlosman}, {Frank}  \& {Begelman}}{{Shlosman}
  et~al.}{1989}]{shlosman89b}
{Shlosman} I.,  {Frank} J.,   {Begelman} M.~C.,  1989, \mn@doi [\nat]
  {10.1038/338045a0}, \href
  {https://ui.adsabs.harvard.edu/abs/1989Natur.338...45S} {338, 45}

\bibitem[\protect\citeauthoryear{{Stone}, {Vasiliev}, {Kesden}, {Rossi},
  {Perets}  \& {Amaro-Seoane}}{{Stone} et~al.}{2020}]{Stone2020}
{Stone} N.~C.,  {Vasiliev} E.,  {Kesden} M.,  {Rossi} E.~M.,  {Perets} H.~B.,
  {Amaro-Seoane} P.,  2020, \mn@doi [\ssr] {10.1007/s11214-020-00651-4}, \href
  {https://ui.adsabs.harvard.edu/abs/2020SSRv..216...35S} {216, 35}

\bibitem[\protect\citeauthoryear{{Takeuchi} \& {Okuzumi}}{{Takeuchi} \&
  {Okuzumi}}{2014}]{takeuchi14}
{Takeuchi} T.,  {Okuzumi} S.,  2014, \mn@doi [\apj]
  {10.1088/0004-637X/797/2/132}, \href
  {https://ui.adsabs.harvard.edu/abs/2014ApJ...797..132T} {797, 132}

\bibitem[\protect\citeauthoryear{{Tozzi} et~al.,}{{Tozzi}
  et~al.}{2022}]{tozzi2022}
{Tozzi} P.,  et~al., 2022, \mn@doi [\aap] {10.1051/0004-6361/202142333}, \href
  {https://ui.adsabs.harvard.edu/abs/2022A&A...662A..54T} {662, A54}

\bibitem[\protect\citeauthoryear{{Venemans} et~al.,}{{Venemans}
  et~al.}{2017}]{venemans2017}
{Venemans} B.~P.,  et~al., 2017, \mn@doi [\apjl] {10.3847/2041-8213/aa943a},
  \href {https://ui.adsabs.harvard.edu/abs/2017ApJ...851L...8V} {851, L8}

\bibitem[\protect\citeauthoryear{{Wise}, {Regan}, {O'Shea}, {Norman}, {Downes}
  \& {Xu}}{{Wise} et~al.}{2019}]{wise2019}
{Wise} J.~H.,  {Regan} J.~A.,  {O'Shea} B.~W.,  {Norman} M.~L.,  {Downes}
  T.~P.,   {Xu} H.,  2019, \mn@doi [\nat] {10.1038/s41586-019-0873-4}, \href
  {https://ui.adsabs.harvard.edu/abs/2019Natur.566...85W} {566, 85}

\bibitem[\protect\citeauthoryear{{Yang} et~al.,}{{Yang}
  et~al.}{2023}]{yang2023}
{Yang} G.,  et~al., 2023, \mn@doi [\apjl] {10.3847/2041-8213/acd639}, \href
  {https://ui.adsabs.harvard.edu/abs/2023ApJ...950L...5Y} {950, L5}

\bibitem[\protect\citeauthoryear{{Yusef-Zadeh} \& {Morris}}{{Yusef-Zadeh} \&
  {Morris}}{1987}]{yusefzadeh87}
{Yusef-Zadeh} F.,  {Morris} M.,  1987, \mn@doi [\apj] {10.1086/165572}, \href
  {https://ui.adsabs.harvard.edu/abs/1987ApJ...320..545Y} {320, 545}

\bibitem[\protect\citeauthoryear{{Yusef-Zadeh} et~al.,}{{Yusef-Zadeh}
  et~al.}{2013}]{yusefzadeh13}
{Yusef-Zadeh} F.,  et~al., 2013, \mn@doi [\apj] {10.1088/0004-637X/762/1/33},
  \href {https://ui.adsabs.harvard.edu/abs/2013ApJ...762...33Y} {762, 33}

\bibitem[\protect\citeauthoryear{{Yusef-Zadeh}, {Arendt}, {Wardle}, {Heywood},
  {Cotton}  \& {Camilo}}{{Yusef-Zadeh} et~al.}{2022}]{yusefzadeh22}
{Yusef-Zadeh} F.,  {Arendt} R.~G.,  {Wardle} M.,  {Heywood} I.,  {Cotton} W.,
  {Camilo} F.,  2022, \mn@doi [\apjl] {10.3847/2041-8213/ac4802}, \href
  {https://ui.adsabs.harvard.edu/abs/2022ApJ...925L..18Y} {925, L18}

\bibitem[\protect\citeauthoryear{{Zrake}}{{Zrake}}{2014}]{zrake14}
{Zrake} J.,  2014, \mn@doi [\apjl] {10.1088/2041-8205/794/2/L26}, \href
  {https://ui.adsabs.harvard.edu/abs/2014ApJ...794L..26Z} {794, L26}

\makeatother
\end{thebibliography}




\end{document}